\documentclass{article}

\usepackage{arxiv}

\usepackage{graphicx}%
\usepackage{multirow}%
\usepackage{amsmath,amssymb,amsfonts}%
\usepackage{amsthm}%
\usepackage{mathrsfs}%
\usepackage[title]{appendix}%
\usepackage{xcolor}%
\usepackage{textcomp}%
\usepackage{manyfoot}%
\usepackage{booktabs}%
\usepackage{algorithm}%
\usepackage{algorithmicx}%
\usepackage{algpseudocode}%
\usepackage{listings}%
\usepackage[authoryear,round]{natbib}
\usepackage{hyperref}
\usepackage{setspace}

\usepackage{bm}
\definecolor{DarkGreen}{HTML}{006400}

\theoremstyle{thmstyleone}%

\theoremstyle{thmstyletwo}%
\newtheorem{remark}{Remark}%

\theoremstyle{thmstylethree}%

\begin{document}

\title{ Structured Covariate-Informed Empirical Orthogonal Functions for Spatio-Temporal Environmental Fields
}

%
%
%

\author{
{\hspace{1mm}Hao-Yun  Huang}
\\
Department of Applied Mathematics,\\
National Dong Hwa University,\\
Hualien, Taiwan\\
\texttt{hhuscout@gms.ndhu.edu.tw}
\And
{\hspace{1mm}ShengLi Tzeng}\thanks{Corresponding author}\\
Department of Applied Mathematics and \\
Graduate Institute of Statistics,\\
National Chung Hsing University,\\
Taichung, Taiwan\\
\texttt{slt.cmu@gmail.com}
}

\date{}

\maketitle

\begin{abstract}

Low-rank representations such as empirical orthogonal function (EOF) decompositions are widely used for analyzing large spatio-temporal environmental fields. However, conventional EOF identifies latent modes solely from covariance structure and does not utilize observed environmental covariates, limiting its ability to incorporate external information into low-rank representations.
This study introduces Structured Covariate-Informed EOF (SCIEOF), a covariate-informed extension of EOF that bridges low-rank dimension reduction and prediction-oriented spatio-temporal modeling. SCIEOF embeds spatial and temporal covariates into the latent bases while incorporating spatio-temporal covariates through an additive component, yielding low-rank representations with latent modes informed by observed covariates. Estimation procedures are developed and evaluated through simulation studies and an application to global near-surface air temperature from the MERRA-2 reanalysis.
Simulation studies demonstrate that incorporating informative covariates improves latent structure recovery and predictive accuracy, particularly when the spatial basis is appropriately specified. The advantage is more pronounced at moderate-to-large sample sizes, while methods with stronger structural assumptions remain competitive when data are limited.  In the MERRA-2 application, SCIEOF achieves competitive or improved predictive performance relative to commonly used methods while providing a compact and physically interpretable low-rank representation.
Overall, SCIEOF provides a flexible and computationally scalable framework for integrating structural covariate information into low-rank spatio-temporal representations, extending EOF toward predictive environmental modeling.

\end{abstract}


\onehalfspacing

\section{Introduction}\label{sec:intro}

Spatio-temporal data arise in a wide range of environmental and geoscientific applications, including climate science, air quality monitoring, hydrology, ecology, and public health. Beyond prediction and description, a central objective is to identify latent structures in observed variability and relate them to known scientific drivers. Modern datasets are therefore often accompanied by rich auxiliary information that can potentially aid interpretation and inference.

A fundamental challenge is how to incorporate these heterogeneous sources of information into a unified low-rank decomposition of a spatio-temporal process. An ideal decomposition should preserve the structured nature of latent factors and enhance interpretability by explicitly linking its components to scientifically meaningful covariates. Such a connection would facilitate scientific interpretation, improve predictive performance, and provide insight into the mechanisms underlying observed spatio-temporal variability.

Empirical Orthogonal Function (EOF) analysis \citep{hotelling1933analysis} remains one of the most widely used tools for spatio-temporal decomposition in environmental sciences. Suppose that measurements are collected at $n$ locations over $T$ time points, resulting in an $n \times T$ observation matrix $\bm{Z}$. EOF analysis represents $\bm{Z}$ using the low-rank decomposition $\bm{Z} \approx \bm{\Phi}\bm{\Lambda}\bm{W}^\prime$, where $\bm{\Phi}$ and $\bm{W}$ contain the leading spatial loading vectors and temporal score vectors, respectively, from a truncated singular value decomposition (SVD). Despite its simplicity and widespread use, standard EOF analysis is entirely data-driven. The resulting spatial loadings are defined only at observed locations and lack direct connections to geographic or environmental characteristics, while temporal scores are similarly unconstrained by external temporal information. Consequently, the extracted modes often prove difficult to interpret scientifically, even when substantial auxiliary information is available.

The challenge becomes particularly apparent when multiple sources of auxiliary information are available. Consider daily average temperature observations collected from a network of weather stations. Such data are often accompanied by several distinct forms of auxiliary information. Spatial covariates, such as distance from the coast, help explain the broad geographic differences associated with continental and maritime climates. Temporal covariates, such as the annual seasonal cycle, describe recurring patterns shared across locations. In addition, spatio-temporal variables, such as local precipitation, vary jointly across space and time and may influence temperature through mechanisms including cloud cover and evapotranspiration. Although this example is used for illustration, similar combinations of spatial, temporal, and spatio-temporal information arise in many scientific applications. Ideally, all of these information sources would contribute to  the decomposed components used to summarize the underlying spatio-temporal variability.

Various extensions have been proposed to address some of these goals. Low-rank spatial models and kriging-based approaches combine dimension reduction with spatial prediction and often incorporate covariates through the mean structure \citep{banerjee2008predictive, cressie2008fixed, tzeng2018resolution}. Additive and semiparametric models allow flexible nonlinear effects of spatial and temporal predictors \citep{lee2011pspline, wood2017gam}, while functional data methods model dependence among spatially indexed temporal trajectories \citep{gromenko2012estimation, kokoszka2019recent}. A particularly influential line of work is the MESA Air framework \citep{fuentes2006transforms, lindstrom2014flexible, sampson2011pragmatic, szpiro2010predicting}, which combines temporal basis functions with spatial regression on geographic covariates. In the temperature example, these approaches can use coastal proximity and related geographic information to explain large-scale spatial variation and improve interpretation of spatial patterns. Related approaches such as EOF regression \citep{thorson2020eof} use EOF-derived components to explain variability in external outcomes. More broadly, existing methods often improve interpretability and predictive performance by incorporating selected sources of auxiliary information.

However, these sources of information typically enter the analysis in fundamentally different ways. For instance, methods like the MESA Air framework successfully guide the spatial components using geographic or environmental covariates, but leave the temporal basis functions entirely data-driven despite the availability of temporal information. Conversely, approaches that use external temporal information can guide temporal factors while leaving the spatial structure fully nonparametric. More importantly, neither perspective naturally accommodates covariates whose effects vary jointly across space and time, such as precipitation in the temperature example. As a result, spatial, temporal, and spatio-temporal covariates are rarely allowed to jointly determine the components of a common low-rank decomposition. Existing supervised extensions of EOF, which integrate auxiliary variables to guide the dimension reduction process, are therefore fundamentally asymmetric in how external information is incorporated into the decomposition.

These observations motivate a more general framework for supervised dimension reduction in spatio-temporal settings, highlighting the need for a low-rank decomposition that avoids asymmetric incorporation of auxiliary information and instead integrates multiple sources within a single coherent factorization. In this article, we propose a Structured Covariate-Informed EOF framework (SCIEOF) that addresses this gap by jointly accounting for both spatial and temporal factors within a unified factorization. The central idea, formalized in Section 2, is to parameterize the spatial loading matrix $\bm{\Phi}$ as lying in the column space of a $n \times p$ matrix $\bm{R}$ of spatial covariates (which may include known geographic attributes, thin-plate splines, or radial basis functions), and the temporal score matrix $\bm{W}$ as lying in the column space of a $T \times q$ matrix $\bm{H}$ of temporal covariates (B-splines, harmonic terms, or observed time-varying predictors such as climate indices). The rank-$L$ approximation to the underlying structure is then estimated by performing an approximate SVD on the doubly-projected matrix $({\bm{R}}'\bm{R})^{-1}{\bm{R}}'\bm{Z}\bm{H}({\bm{H}}'\bm{H})^{-1}$, a $p \times q$ matrix of much smaller dimension than the observed data matrix $\bm{Z}$ when $p \ll n$ and $q \ll T$. The factorized components are thus bilaterally informed by external information. Simultaneously, spatio-temporal covariate matrices $\{\bm{F}_{(k)}; k=1,\ldots,K \}$ are incorporated through an alternating regression step.  This leads to a structured framework in which spatial, temporal, and spatio-temporal information are integrated within a single coherent decomposition, yielding components that are parsimonious and potentially more interpretable when the incorporated covariates carry direct physical or geographic meaning.

The proposed framework yields several important benefits. First, it provides a unified low-rank decomposition in which spatial and temporal covariates directly shape the components, rather than entering only through post hoc modeling steps. Second, the bilateral projection substantially reduces the computational burden for large values of $n$ or $T$. Third, the incorporation of spatial covariates, including spatial basis functions, allows loading functions to be evaluated at unobserved locations, enabling direct spatial interpolation without separate or latent kriging procedures. Fourth, when meaningful spatial or temporal covariates are incorporated into $\bm{R}$ or $\bm{H}$, the estimated loading and score functions can in principle be related to those covariates, offering a potential interpretive advantage not available under standard EOF.

To evaluate the practical performance of the proposed framework, we conducted extensive simulation studies under various spatial and temporal structures. The results demonstrate that SCIEOF substantially outperforms standard EOF analysis, the MESA Air framework, and universal kriging across a range of scenarios. A real data application to global near-surface air temperature illustrates the method's practical advantages.

The remainder of this paper is organized as follows. Section~2 presents the proposed model, estimation and prediction procedures, and its relationship to existing methods. Section~3 evaluates the method through simulation studies under a range of spatial, temporal, and covariate-driven scenarios. Section~4 applies the approach to global temperature data. Section~5 concludes.

\section{Model and Method}\label{sec2}

\subsection{Data Structure and Notation}

Suppose that a univariate spatio-temporal response $Z(\bm{s}, t)$ is observed at $n$ spatial locations $\bm{s}_1, \ldots, \bm{s}_n \in \mathcal{D} \subset \mathbb{R}^d$ and $T$ time points $t = 1, \ldots, T$. The observations are organized into a matrix $\bm{Z} \in \mathbb{R}^{n \times T}$ with $(i, t)$-entry $Z(\bm{s}_i, t)$.

In addition to the response, three sets of covariate matrices are available:
\begin{itemize}
    \item  $\bm{R} \in \mathbb{R}^{n \times p}$: a matrix of spatial-only covariate values, with $i$-th row $\bm{r}(\bm{s}_i)'$;
   \item $\bm{H} \in \mathbb{R}^{T \times q}$: a matrix of temporal-only covariate values, with $t$-th row $\bm{h}(t)'$;
  \item $\{\bm{F}_{(k)} \in \mathbb{R}^{n \times T};\, k = 1, \ldots, K\}$: matrices of spatiotemporal covariates.
\end{itemize}

The columns of $\bm{R}$ may include known geographic attributes (latitude, longitude, elevation, climate zone indicators), polynomial terms in the coordinates, or nonlinear basis functions such as thin-plate splines (\citealp{wood2003thin}) or radial basis functions centered at a set of knot locations across $\mathcal{D}$. The columns of $\bm{H}$ may include seasonal harmonic functions, B-spline bases, or observed temporal covariates such as large-scale climate indices.

\subsection{The SCIEOF Model}

The central modeling assumption is that the signal $\bm{Z}$ can be approximated as

\begin{equation}
\bm{Z} = \bm{Y}+  \bm{E} = \bm{R}\bm{P}\boldsymbol{\Lambda}\bm{Q}'\bm{H}' + \sum_{k=1}^{K} b_k \bm{F}_{(k)} + \bm{E}, \label{eq:SVD_model} 
\end{equation}
where $\bm{Y}$ is the noiseless version of $\bm{Z}$, $\bm{P} \in \mathbb{R}^{p \times L}$ and $\bm{Q} \in \mathbb{R}^{q \times L}$ are matrices of unknown coefficients with orthonormal columns, $\boldsymbol{\Lambda} = \mathrm{diag}(\lambda_1, \ldots, \lambda_L)$ contains nonnegative singular values, $b_k$ are unknown scalar coefficients for the spatiotemporal covariates, and $\bm{E}$ is a residual matrix. 

The $j$-th spatial loading at any location $\bm{s}$ is
\begin{equation}
\phi_j(\bm{s}) = \lambda_j \, \bm{r}(\bm{s})' \bm{p}_j, \label{eq:spatial_loading}
\end{equation}
and the $j$-th temporal score at any time $t$ is
\begin{equation}
w_j(t) = \bm{h}(t)' \bm{q}_j. \label{eq:temporal_loading}
\end{equation}

Equation (\ref{eq:spatial_loading}) allows immediate evaluation at unobserved locations by computing $\bm{r}(\bm{s}_0)$ for any $\bm{s}_0 \notin \{\bm{s}_1, \ldots, \bm{s}_n\}$, and (\ref{eq:temporal_loading}) allows interpolation to new time points within the support of $\bm{h}$.

\begin{remark}
\textbf{ Relationship to classical EOF.} 
 When $\bm{R} = \bm{I}_n$, $\bm{H} = \bm{I}_T$, and $K = 0$, model (\ref{eq:SVD_model}) reduces to the standard rank-$L$ SVD approximation of $\bm{Z}$, i.e., classical EOF analysis. The proposed model thus strictly generalizes EOF: the identity choices correspond to imposing no structure on the loadings or scores, while non-identity choices encode spatial patterns, temporal smoothness, and covariate supervision simultaneously.
\end{remark}

\begin{remark}
\textbf{ Relationship to the MESA Air framework.} MESA Air framework considers models of the form
\begin{equation}
Z(\bm{s}, t) = \beta_0(\bm{s}) + \sum_{j=1}^{m} \beta_j(\bm{s}) m_j(t) + \nu(\bm{s}, t), \label{eq:MESA}
\end{equation}
where the temporal basis functions $m_j(t)$ are estimated by smoothing the leading right singular vectors of a (possibly imputed) data matrix, and the spatially varying coefficients $\beta_j(\bm{s})$ are modeled as functions of land-use geographic covariates. \citet{olives2014reduced} further represent each $\beta_j$-field using thin-plate regression splines to achieve low-rank spatial approximation.
\end{remark}

Our model (\ref{eq:SVD_model}) can be seen as a bilateral generalization of (\ref{eq:MESA}): both the spatial loading functions and the temporal score functions are simultaneously parameterized through their respective basis matrices, and the dominant factor structure is extracted by a single projected SVD rather than by first smoothing the temporal singular vectors and then regressing each location's time series onto them. This bilateral symmetry means that temporal basis functions and temporal covariates can enter the factor structure just as spatial basis functions and spatial covariates do. This is a feature that is absent from the MESA Air framework, where the temporal structure is modeled in a data-driven manner without incorporating covariates.   The MESA Air approach arises as the special case in which $\bm{H} = \bm{I}_T$ (i.e., no temporal basis) and  the rows of $\bm{R}$ contain land-use covariates.

\begin{remark}
\textbf{ Relationship to PDE+.} \citet{lue2023interpretable} estimate a mean structure of the form $\sum_j w_j(t)\phi_j(\bm{s})$ via the enhanced Pairwise Directions Estimation method (PDE+), which treats $\phi_j(\bm{s}) = g_j(\boldsymbol{\theta}_j' \bm{x}(\bm{s}))$ as a nonparametric function of a linear combination of spatial covariates, estimated by minimizing a locally weighted least-squares criterion. Unlike PDE+, our present framework imposes an explicit basis representation on both $\phi_j$ and $w_j$, which makes it a proper reduced-rank regression model with a well-defined matrix objective, facilitates simultaneous treatment of spatial and temporal covariates, and avoids the bandwidth selection and iterative kernel smoothing required by PDE+.
\end{remark}

\subsection{Estimation Algorithm
}\label{subsec2.3}

The estimation objective is to minimize the Frobenius-norm criterion

\begin{equation}
\mathcal{L}(\bm{P}, \bm{Q}, \bm{b}) = \left\|\bm{Z} - \bm{R}\bm{P}\boldsymbol{\Lambda}\bm{Q}'\bm{H}' - \sum_{k=1}^{K} b_k \bm{F}_{(k)}\right\|_F^2, \label{eq:loss}
\end{equation}
subject to $\bm{P}'\bm{P} = \bm{I}_L$ and $\bm{Q}'\bm{Q} = \bm{I}_L$, where $L\leq \min{p,q}$ is a preselected integer. Because (\ref{eq:loss}) is jointly non-convex in $(\bm{P}, \bm{Q}, \bm{b})$, we adopt the following alternating algorithm, which is summarized in Algorithm 1.

\textbf{Step 0 (Initialization).} Set $\hat{\bm{b}}^{(0)} = \bm{0}$ and $\bm{\Delta}^{(0)} = \bm{Z} - \bar{z}\bm{1}_n\bm{1}_T'$, where $\bar{z}$ is the grand mean of $\bm{Z}$.

\textbf{Step 1 (Projected SVD).} Given the current working residual $\bm{\Delta}^{(j)}$, form the $p \times q$ projected matrix

\begin{equation}
\bm{M}^{(j)} = \left(\bm{R}'\bm{R}\right)^{-1}\bm{R}'\bm{\Delta}^{(j)}\bm{H}\left(\bm{H}'\bm{H}\right)^{-1}, \label{eq:MME}
\end{equation}

and compute its rank-$L$ SVD,  $\bm{M}^{(j)} = \hat{\bm{P}}^{(j)}\hat{\boldsymbol{\Lambda}}^{(j)}\hat{\bm{Q}}^{(j)'}$. Set

\begin{equation*}
\hat{\bm{U}}^{(j)} = \bm{R}\hat{\bm{P}}^{(j)}, \qquad \hat{\bm{V}}^{(j)} = \bm{H}\hat{\bm{Q}}^{(j)}. 
\end{equation*}

\textbf{Step 2 (Covariate update).} Regress the residual $\bm{Z} - \hat{\bm{U}}^{(j)}\hat{\boldsymbol{\Lambda}}^{(j)}\hat{\bm{V}}^{(j)'}$ onto $\{\bm{F}_{(k)}; k=1,\ldots,K\}$ (by vectorizing and applying ordinary least squares) to update $\hat{\bm{b}}^{(j+1)}$. Form the new working residual

\begin{equation*}
\bm{\Delta}^{(j+1)} = \bm{Z} - \bar{z}\bm{1}_n\bm{1}_T' - \sum_{k=1}^{K}\hat{b}_k^{(j+1)}\bm{F}_{(k)}.
\end{equation*}

\textbf{Step 3 (Convergence check).} Repeat Steps 1-2 until the relative change in the objective (\ref{eq:loss}) falls below a tolerance $\epsilon$ (e.g., $\epsilon = 10^{-4}$).

The key computational step is (\ref{eq:MME}): instead of performing an $n \times T$ SVD, one solves two ordinary least-squares problems (the left and right projections) and performs a $p \times q$ SVD. When $p \ll n$ and $q \ll T$, as is typically the case when spatial and temporal covariates are used, this represents a substantial reduction in both computational cost and statistical noise. Step 1 is precisely the reduced-rank regression step for the structured factor pair $(\bm{U}, \bm{V})$ conditional on the covariate coefficients, while Step 2 is the ordinary regression step for $\bm{b}$ conditional on the rank-$L$ reconstruction; each step is guaranteed to decrease or leave unchanged the objective (\ref{eq:loss}), ensuring convergence to a stationary point.

The bilateral generalization with non-trivial $\bm{H}$ is the key novel element. When $K = 0$, the algorithm converges in a single step, and the solution is the leading $L$ terms of the SVD of $\bm{M}$. 

\subsection{Selection of \texorpdfstring{$L$}{L} and Basis Dimensions}\label{subsec2.4}
The rank $L$ can be selected by examining the scree plot of singular values of $\bm{M}$ defined in (\ref{eq:MME}), or by cross-validation. The maximal eigenvalue ratio criterion of \citet{luo2009contour} provides a formal rule: $\hat{L} = \arg\max_{j} \hat{\lambda}_j / \hat{\lambda}_{j+1}$. The spatial and temporal dimensions,  $p$ and $q$, should be chosen large enough to span the dominant spatial and temporal patterns,  yet not so large as to cause overfitting; practical guidance suggests $p \approx O(\log_2(n)^2)$ spatial basis functions for $n$ locations, and $q$ equal to the number of harmonic terms or B-spline basis functions needed to represent the dominant temporal cycles.

\subsection{ Prediction}

For a new spatial location $\bm{s}_0$ and time point $t_0$, the predicted signal is

\begin{equation}
\hat{y}(\bm{s}_0, t_0) = \bar{z} + \sum_{j=1}^{L}  \hat{\lambda}_j\hat{u}_j(\bm{s}_0)\hat{v}_j(t_0) + \sum_{k=1}^{K}\hat{b}_k f_{(k)}(\bm{s}_0, t_0), \label{eq:pred}
\end{equation}
where $\hat{u}_j(\bm{s}_0) = \bm{r}(\bm{s}_0)'\hat{\bm{p}}_j$ and $\hat{v}_j(t_0) = \bm{h}(t_0)'\hat{\bm{q}}_j$  are evaluated based on known spatial-only and temporal-only covariate values, or by evaluating the spatial and temporal basis functions at $\bm{s}_0$ and $t_0$. This is a natural and computationally trivial operation, requiring no secondary kriging of the loading functions. Standard EOF, by contrast, provides no mechanism for evaluating the loadings at unobserved locations without an additional modeling step.

\section{Simulation Study}\label{sec:simulation}

We conduct four simulation experiments to evaluate the proposed method and compare it with Standard EOF, SCIEOF variants, MESA Air, and Universal Kriging (UK). Each experiment is designed to isolate one structural feature of the proposed framework:

\begin{itemize}
  \item \textbf{Simulation A}: a rank-2 signal in which the spatial loading functions are nonlinear functions of a \emph{known} purely spatial covariate. This isolates the advantage of encoding geographic structure directly in $\bm{R}$.

  \item \textbf{Simulation B }: a rank-3 signal with one temporal mode driven by a known, abruptly varying covariate $h(t)$ and two smooth, covariate-free temporal modes, highlighting the benefit of encoding the known covariate in $\bm{H}$ instead of estimating it from data.

  \item \textbf{Simulation C }:  a rank-2 deterministic trend driven by smooth, nonlinear spatial loading functions and curved temporal scores, superimposed on a spatially stationary but temporally independent noise field.  This scenario isolates how the error sources from trend misspecification and covariance misspecification  trade off against each other when only one of them favors a given competitor.  

  \item \textbf{Simulation D }: a rank-3 signal with a dominant spatio-temporal covariate and a nonstationary random effect induced by coordinate deformation. This targets the flexibility of the spatial basis to capture locally varying spatial structure that violates the isotropic variogram assumption used in UK, and exposes MESA Air's two-step estimation bias under nonstationarity. 
\end{itemize}

All simulations draw $n$ spatial locations $\bm{s}_i$ using simple random sampling within $[-1,1]^2$ and $T$ equally spaced time points, with $(n,T) \in \{(100,20),\,(500,50),\,(1000,100),\,(2000,200),\,(5000,200)\}$. Totally, 20\% of sites and 20\% of time points are completely held out for evaluation, and all results are averaged over 100 replications.

\subsection{Compared Methods}

We compare the following methods across all four simulations.

\textbf{Standard EOF.} 
The leading $L$ terms of the SVD of the demeaned (and covariate-adjusted) data matrix, with spatial loadings evaluated only at the observed locations. Prediction at held-out locations uses thin-plate spline interpolation of the estimated loadings \citep{perry2008generation}, and prediction at held-out time points uses B-spline extrapolation of the estimated scores \citep{mason2002comparison}.

\textbf{SCIEOF-R2.} Our proposed method with $\bm{R}$ comprising the observed $p_0$ spatial-only covariates augmented by Multi-Resolution Thin-plate Splines (MRTS) with $\min(200, n-5)$ basis functions \citep{tzeng2018resolution}, and $\bm{H}$ comprising the observed $q_0$ temporal-only covariates augmented by a B-spline basis with $6$ functions. Hence, $p = p_0 + \min(200, n-5)$, and $ q = q_0 + 6.$  The rank $L$ is selected by internal cross-validation: $20\%$ of training sites are held out, $L \in \{2,3,4,5,6,7,8\}$ is evaluated, and the value achieving the lowest validation RMSE is selected.

\textbf{SCIEOF-RBF.} A baseline SCIEOF variant in which $\bm{R}$ comprising the observed $p_0$ spatial-only covariates augmented by five spatial polynomial terms $(1, s_1, s_2, s_1^2, s_2^2)$ and  $16$ radial basis functions (RBF) centered on a regular grid over $[-1,1]^2$, i.e., $p=p_0+21$ by replaceing the MRTS basis with a more conventional spatial basis. $\bm{H}$ and the rank-selection procedure are unchanged from SCIEOF-R2.

\textbf{MESA Air.} We follow the modeling framework of \citet{lindstrom2014flexible}, in which the mean process is represented as a linear combination of fixed spatio-temporal covariates and temporal basis functions extracted by SVD from the covariate-adjusted training data (extended to unobserved time points by cubic-spline interpolation), whose coefficients vary smoothly over space: each coefficient field is modeled by a land-use-regression mean plus a spatially correlated residual and predicted at new sites by universal kriging, with a sparse Gaussian-process residual capturing remaining space-time variation. The temporal basis includes the constant function $m_0 \equiv 1$, whose coefficient $\beta_0(\bm{s})$ is the spatial intercept field. For each coefficient field, the exponential-variogram sill and range are estimated by restricted maximum likelihood with a fixed nugget set to $5\%$ of the field variance.																							  

\textbf{Universal Kriging (UK).} When the spatio-temporal dataset is of moderate size, we employ  parametric kriging with a product-sum covariance model:
\[
C(d_s, d_t) =  C_s(d_s; \xi_1) +  C_t(d_t; \xi_2) + \alpha C_s(d_s; \xi_1) C_t(d_t; \xi_2),
\]
with exponential forms for $C_s$ and $C_t$. For larger datasets, covariance parameters are estimated from subsamples of observations using a full Gaussian process likelihood, while prediction is performed using a nearest-neighbor local kriging approximation based on a small number of nearby spatio-temporal observations that was proposed by \citet{datta2016nonseparable}.

\subsection{Performance Measures}\label{sec:measures}

The competing methods are evaluated using the following criteria, computed on the held-out test set and averaged over the 100 replications:

\begin{itemize}
\item \textbf{RPMSE} (Root Predictive Mean Squared Error):
\[
\text{RPMSE} = \left( \frac{1}{n_\text{test} T} \sum_{t=1}^{T} \sum_{i=1}^{n_\text{test}} \bigl( y (\bm{s}_i, t) - \hat{y}( \bm{s}_i, t) \bigr)^2 \right)^{1/2}.
\]

\item \textbf{CRPS} (Continuous Ranked Probability Score):
\[
\text{CRPS}(F, z) = -\int_{-\infty}^{\infty} \bigl\{ F(v) - \mathbb{I}(v \geq z) \bigr\}^2 \, dv,
\]
where $F(\cdot)$ is the cumulative distribution function of the predictive distribution, $z$ is the observed value, and $\mathbb{I}(\cdot)$ is the indicator function \citep{gneiting2007strictly}.

\item \textbf{Subspace alignment of the low-rank structures:}  
We also evaluate subspace recovery by comparing the estimated spatial and temporal decomposition spaces with the true generating subspaces.

For each replication, we subtract the estimated covariate contribution $\sum \hat{\beta}_k \bm{f}_{(k)}(s,t)$  from the predicted field 
$\hat{z}(s_i, t)$  at all $n$ locations and $T$ time points. We then apply SVD to the residual matrix to extract the leading $ L = \delta$ post-hoc components
\[
\hat{\mathbf{\Phi}}  \in \mathbb{R}^{n \times \delta}, \quad 
\hat{\mathbf{W}} \in \mathbb{R}^{T \times \delta},
\]
where $\delta$ represents the true dimension.

For the proposed SCIEOF and Standard EOF,  $\hat{\mathbf{\Phi}}$  and  $\hat{\mathbf{W}}$  are taken directly from the fitted model. For MESA Air and UK, this post-hoc SVD provides a comparable low-rank representation of the implicit spatial and temporal structure in their predictions \citep{morfin2012spatio, decaens2001spatio}. Sign ambiguity is resolved by Procrustes rotation.

Let 
\[
\mathbf{\Phi}_{\mathrm{true}} \in \mathbb{R}^{n \times \delta}, \quad 
\mathbf{W}_{\mathrm{true}} \in \mathbb{R}^{T \times \delta}
\]
be the true spatial and temporal matrices evaluated at the same points. Alignment between true and estimated subspaces is quantified by the principal angles $\theta_k = \arccos(\sigma_k), $
where $\sigma_k$ are the singular values of
$\mathbf{\Phi}_{\mathrm{true}}^\top \hat{\mathbf{\Phi}}$
after column orthonormalization. The alignment is summarized by the Grassmann distance:
\[
d_{\mathrm{Gr}}(\hat{\mathbf{\Phi}}, \mathbf{\Phi}_{\mathrm{true}}) 
= \left( \sum_{k=1}^{\delta} \theta_k^2 \right)^{1/2}.
\]
The same formula is applied to the temporal matrices. Smaller values indicate better subspace recovery.

\end{itemize}

Smaller values of all these indices indicate better performance. RIMSE and RPMSE assess point-prediction accuracy of the mean field, CRPS evaluates the quality and calibration of predictive distributions, and Grassmann distances measure structural subspace recovery.

\subsection{Simulation A: Spatial Covariate in Loading Functions}\label{sec:simA}

This simulation isolates the advantage of encoding a known geographic attribute directly in $\bm{R}$. The key feature is that the true spatial loading functions are \emph{nonlinear} functions of a purely spatial covariate $g(\bm{s})$ that is observed by the analyst. 
Let $x_1$ and $x_2$ be the two components of a location $\bm{s} \in \mathbb{R}^2$, i.e., $\bm{s}=(x_1,x_2)^\prime$.

Define the purely spatial covariate
\[
  r(\bm{s}) = x_1 + 0.5 x_2^2,
\]
representing a known geographic attribute (e.g.\ a terrain elevation proxy). The signal are generated from
\begin{equation}\label{eq:dgpA}
  Y(\bm{s}_i, t) = \sum_{j=1}^{2} w_j(t)\,\phi_j(\bm{s}_i) + b_s\,r(\bm{s}_i) + \nu(\bm{s}_i,t),
  \quad i = 1, \ldots, n,\ t = 1, \ldots, T,
\end{equation}
where the loading functions are nonlinear in $r$:
\[
  \phi_1(\bm{s}) = \sin\!\bigl(\pi\, r(\bm{s})\bigr), \qquad
  \phi_2(\bm{s}) = r(\bm{s})^2 - \mathbb{E}[r(\bm{s})^2],
\]
with bounded temporal scores
\[
  w_1(t) = 10\sin(2\pi t/T), \qquad w_2(t) = 5\cos(4\pi t/T),
\]
a pure-spatial intercept coefficient $b_s = 3$, and a stationary process $\nu(\bm{s},t)$ with $\mathrm{cov}(\nu(\bm{s},t),\nu(\bm{s}^*,t^*)) = \exp(-\|\bm{s}-\bm{s}^*\|/0.5)$ if $t=t^*$ and $0$ otherwise. Observation noise is $\varepsilon(\bm{s}_i,t) \overset{\mathrm{i.i.d.}}{\sim} N(0,1)$.

\begin{table}[ht]
\centering
\caption{Simulation A: RPMSE (all test), mean (sd) over 100 replications. Best per row is bold.}
\label{tab:simA_rpmse}
\begin{tabular}{lccccc}
\toprule
Setting & EOF & SCIEOF-RBF & SCIEOF-R2 & MESA Air & UK \\
\midrule
$n = 100$, $T = 20$ & 1.83 (0.61) & 1.81 (0.78) & 6.98 (7.58) & 1.91 (0.38) & \textbf{1.60} (0.24) \\
$n = 500$, $T = 50$ & 1.57 (0.14) & 1.52 (0.33) & 1.74 (0.39) & 1.28 (0.13) & \textbf{1.11} (0.07) \\
$n = 1000$, $T = 100$ & 1.51 (0.05) & 1.88 (0.24) & 1.89 (0.16) & 1.15 (0.08) & \textbf{1.07} (0.05) \\
$n = 2000$, $T = 200$ & 1.50 (0.03) & \textbf{1.02} (0.02) & \textbf{1.02} (0.02) & 1.10 (0.04) & 1.07 (0.04) \\
$n = 5000$, $T = 200$ & 1.49 (0.03) & \textbf{1.01} (0.02) & \textbf{1.01} (0.02) & 1.09 (0.04) & 1.12 (0.05) \\
\bottomrule
\end{tabular}
\end{table}

\begin{table}[!ht]
\centering
\caption{Simulation A: CRPS and Grassmann distances, mean~(sd) over 100 replications. Best per row is bold.}
\label{tab:simA_metrics}
\small
\setlength{\tabcolsep}{4pt}
\begin{tabular}{l l rrrrr}
\toprule
Setting $(n,T)$ & Metric & EOF & SCIEOF-RBF & SCIEOF-R2 & MESA Air & UK  \\
\midrule
\multirow{3}{*}{$(100, 20)$}
  & CRPS         & 0.97(0.24) & \textbf{0.96}(0.33) & 2.52(1.71) & 1.11(0.14) & 1.12(0.13) \\
  & $d_G^{(s)}$  & \textbf{0.45}(0.08) & 0.49(0.13) & 1.06(0.40) & 0.52(0.09) & 0.51(0.10) \\
  & $d_G^{(t)}$  & 1.02(0.06) & 1.04(0.08) & 1.17(0.26) & \textbf{0.85}(0.06) & 0.92(0.07) \\[4pt]
\multirow{3}{*}{$(500, 50)$}
  & CRPS         & 0.86(0.06) & 0.86(0.18) & 0.98(0.23) & 0.85(0.05) & \textbf{0.78}(0.04) \\
  & $d_G^{(s)}$  & 0.41(0.03) & \textbf{0.27}(0.09) & 0.31(0.24) & 0.43(0.03) & 0.42(0.03) \\
  & $d_G^{(t)}$  & 1.04(0.03) & 0.57(0.19) & \textbf{0.54}(0.34) & 0.82(0.03) & 0.83(0.03) \\[4pt]
\multirow{3}{*}{$(1000, 100)$}
  & CRPS         & 0.84(0.02) & 1.07(0.14) & 1.08(0.09) & 0.84(0.04) & \textbf{0.74}(0.03) \\
  & $d_G^{(s)}$  & 0.40(0.03) & 0.19(0.06) & \textbf{0.14}(0.02) & 0.41(0.03) & 0.40(0.02) \\
  & $d_G^{(t)}$  & 1.06(0.02) & 0.45(0.16) & \textbf{0.20}(0.10) & 0.83(0.02) & 0.83(0.02) \\[4pt]
\multirow{3}{*}{$(2000, 200)$}
  & CRPS         & 0.83(0.02) & \textbf{0.59}(0.01) & \textbf{0.59}(0.01) & 0.87(0.02) & 0.73(0.02) \\
  & $d_G^{(s)}$  & 0.40(0.02) & \textbf{0.39}(0.02) & \textbf{0.39}(0.02) & 0.40(0.02) & 0.40(0.02) \\
  & $d_G^{(t)}$  & 1.06(0.02) & \textbf{0.81}(0.01) & \textbf{0.81}(0.01) & 0.82(0.01) & 0.83(0.01) \\[4pt]
\multirow{3}{*}{$(5000, 200)$}
  & CRPS         & 0.83(0.02) & \textbf{0.59}(0.01) & \textbf{0.59}(0.01) & 0.89(0.06) & 0.75(0.03) \\
  & $d_G^{(s)}$  & 0.40(0.01) & \textbf{0.39}(0.01) & \textbf{0.39}(0.01) & \textbf{0.39}(0.01) & 0.40(0.01) \\
  & $d_G^{(t)}$  & 1.06(0.02) & \textbf{0.82}(0.01) & \textbf{0.82}(0.01) & \textbf{0.82}(0.01) & 0.83(0.01) \\
\bottomrule
\end{tabular}
\end{table}

Tables~\ref{tab:simA_rpmse}--\ref{tab:simA_metrics} reveal a clear transition in performance as the sample size increases.  UK achieves the lowest RPMSE in the smallest setting, whereas SCIEOF-R2 is unstable due to the high-dimensional MRTS basis relative to the available data.

As $n$ increases, SCIEOF-RBF and SCIEOF-R2  increasingly become competitive, and from n = 2000 onward they achieve the lowest prediction errors, while also providing the most accurate subspace recovery.  In contrast, standard EOF shows little gain with increasing sample size, highlighting the limitation of unsupervised SVD for recovering covariate-driven loading structures.

\subsection{Simulation B: Abruptly Varying Temporal Covariate}\label{sec:simB}

This simulation isolates the advantage of the bilateral structure when a known temporal covariate $h(t)$ varies abruptly over time. MESA Air cannot incorporate $h(t)$ directly into its temporal basis functions $m_j(t)$, which are estimated from data under a global smoothness  assumption. When the true temporal forcing is not smooth, MESA Air's smoothing step systematically attenuates the abrupt variation, producing a bias that grows with the amplitude of $h(t)$. UK has no mechanism for incorporating $h(t)$ into the temporal mean structure and must absorb it into the spatial drift or kriging residuals.

Let $h(t)$ be a piecewise-constant temporal covariate with 3--5 randomly generated segments, where each segment has a minimum length of 3 time points and the segment boundaries are randomly selected, taking values $\pm 5$ in  alternating segments (a step-function climate index). With $\bm{s}=(x_1,x_2)^\prime$,  the signals are generated by

\begin{equation}\label{eq:dgp4}
  Y(\bm{s}_i, t) = h(t)\,\phi_0(\bm{s}_i) + \sum_{j=1}^{2} w_j(t)\,\phi_j(\bm{s}_i) + \nu(\bm{s}_i,t),
\end{equation}
where
\[
  \phi_0(\bm{s}) = 3 x_1 x_2 - \frac{3}{4} (x_1-x_2) 
\]
is the spatially varying sensitivity to $h(t)$ (e.g. regions farther from the center, particularly near the corners, are most responsive to the abrupt forcing).  The remaining low-rank signal uses $\phi_1(\bm{s}) = x_1^2 - x_2^2$  and $\phi_2(\bm{s}) = x_1^3 - 3x_1 x_2^2$ for spatial loadings, while temporal scores are $5\bigl(16\tau^5 - 20\tau^3 + 5\tau\bigr)$ and $3\bigl(4\tau^3 - 3\tau\bigr)$  with $ \tau=2 t/T-1$.  The process $\nu(\cdot,\cdot)$ is stationary exponential with $\sigma_\nu=0.5$ and range $\rho=1$, and observation noise is $\varepsilon \overset{\mathrm{i.i.d.}}{\sim} N(0,1)$.

\begin{table}[ht]
\centering
\caption{Simulation~B: RPMSE (all test), mean (sd) over 100 replications. Best per row is bold.}
\label{tab:simB_rpmse}

\begin{tabular}{lccccc}
\toprule
Setting & EOF & SCIEOF-RBF & SCIEOF-R2 & MESA Air & UK \\
\midrule
$n = 100$, $T = 20$ & 1.79 (0.60) & 2.21 (1.12) & 2.40 (1.34) & 1.63 (1.79) & \textbf{1.18} (0.15) \\
$n = 500$, $T = 50$ & 1.74 (0.37) & 2.26 (0.49) & 1.49 (0.36) & 1.28 (0.18) & \textbf{1.09} (0.11) \\
$n = 1000$, $T = 100$ & 1.92 (0.28) & 2.66 (0.42) & 1.28 (0.23) & 1.27 (0.36) & \textbf{1.07} (0.06) \\
$n = 2000$, $T = 200$ & 2.15 (0.21) & 1.91 (0.29) & \textbf{1.05} (0.05) & 1.18 (0.06) & 1.06 (0.04) \\
$n = 5000$, $T = 200$ & 2.16 (0.21) & 2.08 (0.27) & \textbf{1.04} (0.05) & 1.27 (0.41) & 1.09 (0.05) \\
\bottomrule
\end{tabular}
\end{table}

\begin{table}[!ht]
\centering
\caption{Simulation B: CRPS and Grassmann distances, mean~(sd) over 100 replications. Best per row is bold.}
\label{tab:simB_metrics}
\small
\setlength{\tabcolsep}{4pt}
\begin{tabular}{l l rrrrr}
\toprule
Setting $(n,T)$ & Metric & EOF & SCIEOF-RBF & SCIEOF-R2 & MESA Air & UK  \\
\midrule
\multirow{3}{*}{$(100, 20)$}
  & CRPS         & 0.98(0.27) & 1.20(0.53) & 1.31(0.63) & 0.91(0.63) & \textbf{0.75}(0.08) \\
  & $d_G^{(s)}$  & 0.70(0.27) & 0.86(0.24) & 1.19(0.32) & 0.70(0.27) & \textbf{0.65}(0.27) \\
  & $d_G^{(t)}$  & 1.23(0.27) & \textbf{1.21}(0.30) & 1.39(0.29) & \textbf{1.21}(0.32) & 1.24(0.32) \\[4pt]
\multirow{3}{*}{$(500, 50)$}
  & CRPS         & 0.96(0.17) & 1.28(0.28) & 0.83(0.20) & 0.74(0.07) & \textbf{0.70}(0.05) \\
  & $d_G^{(s)}$  & 0.50(0.25) & 0.86(0.20) & 0.61(0.33) & 0.47(0.24) & \textbf{0.47}(0.24) \\
  & $d_G^{(t)}$  & 1.19(0.27) & \textbf{1.03}(0.35) & 1.15(0.33) & 1.17(0.29) & 1.15(0.31) \\[4pt]
\multirow{3}{*}{$(1000, 100)$}
  & CRPS         & 1.06(0.15) & 1.50(0.24) & 0.73(0.13) & 0.73(0.08) & \textbf{0.69}(0.03) \\
  & $d_G^{(s)}$  & 0.39(0.17) & 0.93(0.20) & 0.39(0.17) & 0.37(0.17) & \textbf{0.36}(0.16) \\
  & $d_G^{(t)}$  & 1.11(0.24) & 1.41(0.29) & \textbf{1.06}(0.27) & 1.12(0.27) & 1.10(0.26) \\[4pt]
\multirow{3}{*}{$(2000, 200)$}
  & CRPS         & 1.21(0.14) & 1.08(0.16) & \textbf{0.61}(0.02) & 0.72(0.02) & 0.68(0.02) \\
  & $d_G^{(s)}$  & 0.29(0.10) & 0.74(0.06) & \textbf{0.27}(0.10) & 0.29(0.11) & 0.29(0.11) \\
  & $d_G^{(t)}$  & 1.09(0.26) & \textbf{0.65}(0.26) & 1.13(0.26) & 1.17(0.24) & 1.16(0.25) \\[4pt]
\multirow{3}{*}{$(5000, 200)$}
  & CRPS         & 1.21(0.13) & 1.17(0.16) & \textbf{0.60}(0.02) & 0.74(0.07) & 0.67(0.02) \\
  & $d_G^{(s)}$  & 0.29(0.11) & 0.76(0.04) & \textbf{0.25}(0.09) & 0.29(0.12) & 0.27(0.10) \\
  & $d_G^{(t)}$  & 1.10(0.25) & \textbf{0.58}(0.23) & 1.16(0.26) & 1.21(0.25) & 1.18(0.25) \\
\bottomrule
\end{tabular}
\end{table}

Tables \ref{tab:simB_rpmse}--\ref{tab:simB_metrics} show a markedly different pattern from the earlier simulations. UK attains the lowest RPMSE and CRPS in the smallest setting ($n=100$), and SCIEOF-R2 is again unstable at this sample size, reflecting the difficulty of estimating many MRTS basis coefficients with limited data. As $n$ grows, SCIEOF-R2 becomes the clear leader in both point prediction and probabilistic accuracy, achieving the lowest RPMSE and CRPS from $n=2000$ onward and continuing to improve as $n$ increases, while UK's advantage over SCIEOF-R2 disappears once sufficient data are available. SCIEOF-RBF, by contrast, does not achieve competitive performance at any sample size, suggesting that an arbitrary low-dimensional RBF basis may fail to capture challenging spatial structures. This highlights the importance of using an appropriate spatial basis, such as MRTS, for stable and accurate spatial representation.

The spatial and temporal subspace recovery metrics provide further insight. For $d_G(s)$, SCIEOF-RBF performs worst  for $n \ge 500$, also indicating that its spatial basis is insufficient to represent the true loading functions. In contrast, SCIEOF-R2 achieves substantially better spatial recovery due to the flexibility of the MRTS basis. The temporal loadings are more challenging to recover for  most methods: $d_G(t)$ shows limited improvement, and in some cases worsens, as $n$ and $T$ increase, with SCIEOF-RBF being the notable exception. 
This indicates that good temporal subspace recovery does not necessarily coincide with good spatial subspace recovery.
 Overall, the results highlight a trade-off between spatial flexibility and temporal subspace estimation, with SCIEOF-R2 providing superior spatial representation and overall predictive performance.

\subsection{Simulation C: Strong Trend with Stationary Spatial Field} \label{sec:simC}
This simulation is adapted from Example~3.1 of \citet{lue2023interpretable}.
 It splits the advantage across two \emph{different} components of the data-generating process and lets them blend together. The random-effect part $\nu(\bm{s},t)$ is exactly the stationary isotropic exponential field assumed by UK's covariance model, and it is uncorrelated across time, so UK's parametric form for the noise is correctly specified. The mean part, however, is a deterministic rank-2 signal $\sum_j w_j(t)\phi_j(\bm{s})$ with nonlinear spatial loadings and time-varying scores. The temporal coordinate enters only through this deterministic mean structure and not through the covariance of the random effect. This simulation therefore tests how the methods divide their error between mean misspecification and covariance misspecification, rather than testing covariance estimation alone.  With $\bm{s}=(x_1,x_2)^\prime$,  the signal is
\begin{equation}\label{eq:dgpC}
  Y(\bm{s}_i, t) = \sum_{j=1}^2 w_j(t)\,\phi_j(\bm{s}_i) + b_1\,f_1(\bm{s}_i,t) + \nu(\bm{s}_i,t),
  \quad i = 1, \ldots, n,\ t = 1, \ldots, T,
\end{equation}
with
\begin{alignat*}{2}
  \phi_1(\bm{s}) &= \cos\!\bigl(0.5\pi\|\bm{s} - (-0.5,-0.5)'\|^2\bigr), &\qquad
  \phi_2(\bm{s}) &= \sin\!\bigl(0.5\pi\|\bm{s} - (0.5,0.5)'\|^2\bigr), \\
  w_1(t) &= (0.5t - 5)^2, &\qquad
  w_2(t) &= 5\sin(0.1\pi t),
\end{alignat*}
the spatio-temporal covariate $f_1(\bm{s},t) = \exp(x_1 \log t + x_2)$ with $b_1 = 0.3$, a zero-mean random effect $\nu(\bm{s},t)$ with covariance $\mathrm{cov}(\nu(\bm{s},t),\nu(\bm{s}^*,t^*)) = \exp(-0.5\|\bm{s} - \bm{s}^*\|)$ if $t = t^*$ and $0$ otherwise, and observation noise $\varepsilon(\bm{s}_i, t) \overset{\mathrm{i.i.d.}}{\sim} N(0, 0.25)$.

Tables~\ref{tab:simC_rpmse} and~\ref{tab:simC_metrics} report results over 100 replications.

\begin{table}[ht]
\centering
\caption{Simulation C: RPMSE (all test), mean (sd) over 100 replications. Best per row is bold.}
\label{tab:simC_rpmse}
\begin{tabular}{lccccc}
\toprule
Setting & EOF & SCIEOF-RBF & SCIEOF-R2 & MESA Air & UK \\
\midrule
$n = 100$, $T = 20$ & 1.88 (0.69) & \textbf{1.48} (0.47) & 1.51 (0.44) & 2.95 (0.84) & 2.35 (1.31) \\
$n = 500$, $T = 50$ & 3.68 (0.69) & 4.82 (0.94) & \textbf{2.06} (0.93) & 10.11 (2.61) & 9.01 (7.52) \\
$n = 1000$, $T = 100$ & 11.72 (2.78) & 14.41 (1.69) & \textbf{6.33} (2.37) & 25.67(12.24) & 40.12 (25.76) \\
$n = 2000$, $T = 200$ & 44.99 (9.35) & 47.34 (3.55) & \textbf{24.38} (6.72) & 51.38(14.06) & 185.88 (92.13) \\
$n = 5000$, $T = 200$ & 47.23 (6.47) & 27.22 (1.37) & 23.25 (4.21) & \textbf{21.34}(6.58) & 289.32 (80.97) \\
\bottomrule
\end{tabular}
\end{table}

\begin{table}[!ht]
\centering
\caption{Simulation C: CRPS and Grassmann distances, mean~(sd) over 100 replications. Best per row is bold.}
\label{tab:simC_metrics}
\small
\setlength{\tabcolsep}{4pt}
\begin{tabular}{l l rrrrr}
\toprule
Setting $(n,T)$ & Metric & EOF & SCIEOF-RBF & SCIEOF-R2 & MESA Air & UK  \\
\midrule
\multirow{3}{*}{$(100, 20)$}
  & CRPS         & 0.99(0.29) & \textbf{0.79}(0.22) & 0.84(0.23) & 1.41(0.24) & 1.38(0.56) \\
  & $d_G^{(s)}$  & 0.50(0.17) & 0.44(0.11) & \textbf{0.43}(0.13) & 0.46(0.11) & 0.44(0.17) \\
  & $d_G^{(t)}$  & 0.29(0.19) & 0.24(0.12) & \textbf{0.20}(0.14) & 0.34(0.15) & 0.34(0.16) \\[4pt]
\multirow{3}{*}{$(500, 50)$}
  & CRPS         & 1.92(0.26) & 2.17(0.29) & \textbf{0.82}(0.13) & 10.23(0.62) & 3.72(2.31) \\
  & $d_G^{(s)}$  & \textbf{1.29}(0.16) & 1.30(0.17) & 1.30(0.17) & 1.30(0.17) & 1.30(0.17) \\
  & $d_G^{(t)}$  & 1.49(0.05) & \textbf{1.44}(0.04) & \textbf{1.44}(0.04) & \textbf{1.44}(0.04) & \textbf{1.44}(0.04) \\[4pt]
\multirow{3}{*}{$(1000, 100)$}
  & CRPS         & 5.52(1.24) & 6.47(0.55) & \textbf{2.04}(0.22) & 60.06(9.50) & 16.49(7.06) \\
  & $d_G^{(s)}$  & \textbf{1.25}(0.14) & \textbf{1.25}(0.15) & \textbf{1.25}(0.15) & 1.26(0.15) & 1.26(0.15) \\
  & $d_G^{(t)}$  & \textbf{1.48}(0.01) & 1.50(0.01) & 1.50(0.01) & 1.50(0.01) & 1.50(0.01) \\[4pt]
\multirow{3}{*}{$(2000, 200)$}
  & CRPS         & 20.91(4.27) & 21.60(1.08) & \textbf{8.74}(0.74) & 298.38(48.76) & 75.14(26.98) \\
  & $d_G^{(s)}$  & \textbf{1.18}(0.11) & \textbf{1.18}(0.12) & \textbf{1.18}(0.12) & \textbf{1.18}(0.12) & 1.18(0.12) \\
  & $d_G^{(t)}$  & \textbf{1.52}(0.00) & \textbf{1.52}(0.00) & \textbf{1.52}(0.00) & 1.53(0.00) & 1.53(0.01) \\[4pt]
\multirow{3}{*}{$(5000, 200)$}
  & CRPS         & 22.02(3.26) & 12.95(0.51) & \textbf{9.14}(0.64) & 320.39(7.71) & 110.68(29.30) \\
  & $d_G^{(s)}$  & \textbf{1.17}(0.07) & \textbf{1.17}(0.07) & \textbf{1.17}(0.07) & \textbf{1.17}(0.07) & \textbf{1.17}(0.08) \\
  & $d_G^{(t)}$  & \textbf{1.52}(0.00) & \textbf{1.52}(0.00) & \textbf{1.52}(0.00) & 1.53(0.00) & 1.53(0.01) \\
\bottomrule
\end{tabular}
\end{table}

Table~\ref{tab:simC_rpmse} shows that correctly specifying the residual covariance is not sufficient when the mean structure is nonlinear and low-rank. Although UK uses the true covariance family, it is never the best-performing method and becomes the worst from $n=1000$ onward, with RPMSE increasing sharply as sample size grows. In contrast, SCIEOF-R2 is the best or tied-best method at every sample size from $n=500$ onward, indicating that directly modeling the low-rank mean structure is more important than correctly specifying the covariance in this setting.    MESA Air generally outperforms UK but performs worse than SCIEOF-R2. The exception is at the largest sample size, where MESA Air achieves the lowest RPMSE, slightly outperforming SCIEOF-R2.

The CRPS results in Table~\ref{tab:simC_metrics}  further highlight this contrast, with SCIEOF-R2 producing the most accurate predictive distributions and MESA Air  showing substantially poorer predictive distributions at larger sample sizes. By contrast, Grassmann distances are nearly identical across methods for $n\ge500$, suggesting that all methods recover similar low-dimensional subspaces even when their predictive accuracy differs substantially.  In particular, at $n=5000$, MESA Air achieves the lowest RPMSE but has the highest CRPS.  Performance differences   appear to  arise primarily from how accurately they estimate the associated scores and overall trend magnitude rather than from the geometric structure of the subspace. Together, these results indicate that when the dominant signal is a nonlinear low-rank mean structure, accurate mean modeling   plays a greater role in  prediction performance than correct specification of the residual covariance.

\subsection{Simulation D: Spatial Nonstationarity via Deformation}\label{sec:simD}

This simulation targets the MRTS basis's ability to handle nonstationary spatial loading functions. Simulation~4 has a rank-3 smooth trend and a strong spatio-temporal covariate, similar to those in Simulation~1, but replaces the stationary $\nu(\cdot,\cdot)$ with a nonstationary Gaussian process obtained through the coordinate deformation approach of Sampson \& Guttorp~(1992). Here let $\bm{s}=(x_1, x_2)^\prime$ again.  Define
\[
  \psi(\bm{s}) = \bigl(\mathrm{sign}(x_1)|x_1|^{a},\ \mathrm{sign}(x_2)|x_2|^{b}\bigr),
  \qquad a = 2.5,\ b = 0.4,
\]
so the two coordinate axes are warped by \emph{different} powers: $x_1$ is compressed near the origin while $x_2$ is expanded. In the deformed space the random effect is stationary exponential,
\[
  \mathrm{cov}\bigl(\nu(\bm{s},t),\, \nu(\bm{s}^*,t^*)\bigr)
  = \sigma_\nu^2 \, \exp\!\bigl(-\|\psi(\bm{s}) - \psi(\bm{s}^*)\|/\rho\bigr) \cdot \mathbf{1}_{t=t^*},
  \qquad \sigma_\nu = 0.7,\ \rho = 0.1,
\]
which induces anisotropic, location-dependent covariance in the original coordinates. 

The mean structure is a rank-3 signal plus a dominant spatio-temporal covariate:
\begin{equation}\label{eq:dgpD}
  Y(\bm{s}_i, t) = \sum_{j=1}^{3} w_j(t)\,\phi_j(\bm{s}_i) + b_1\,f_1(\bm{s}_i,t) + \nu(\bm{s}_i,t),
\end{equation}
with
\begin{alignat*}{2}
  \phi_j(\bm{s})
  &= \exp\!\bigl(-2\|\bm{s}-\mathbf{c}_j\|^2\bigr),
  &\quad
  \mathbf{c}_1 &= (-0.5,-0.5)',\quad
  \mathbf{c}_2=(0.5,0.5)',\quad
  \mathbf{c}_3=(-0.5,0.5)',\\[1mm]
  w_1(t)
  &= 10\sin(2\pi t/T),
  &\quad
  w_2(t) &= 5\cos(4\pi t/T),\quad
  w_3(t)=3\sin(6\pi t/T).
\end{alignat*}
and the spatio-temporal covariate
\[
  f_1(\bm{s},t) = 5\sin(\pi s_1 t/T)\cos(\pi s_2) + 3(s_1^2 - s_2^2)\cos(2\pi t/T),
\]
having the coefficient $b_1 = 5$. Observation noise is $\varepsilon \overset{\mathrm{i.i.d.}}{\sim} N(0,1)$.

The coordinate deformation  induces a nonstationary, anisotropic spatial covariance in $\nu(\mathbf{s}, t)$, causing spatial dependence to vary rapidly along the $s_1$ direction near the origin (compressed axis) and slowly along $s_2$ (expanded axis). UK's isotropic variogram cannot represent this anisotropy and is systematically misspecified. MESA Air's global smoothing of the temporal singular vectors may  average out the locally rapid variation in the spatial loadings, limiting its ability to capture the locally varying spatial structure.   The MRTS multi-resolution basis in SCIEOF-R2 naturally captures different spatial frequencies at different locations, adapting to the locally varying structure without any additional model specification.

\begin{table}[ht]
\centering
\caption{Simulation~D: RPMSE (all test), mean (sd) over 100 replications. Best per row is bold.}
\label{tab:simD_rpmse}
\begin{tabular}{lccccc}
\toprule
Setting & EOF & SCIEOF-RBF & SCIEOF-R2 & MESA Air & UK \\
\midrule
$n = 100$, $T = 20$ & 5.46 (3.16) & 5.72 (3.37) & 6.25 (3.32) & 2.56 (0.55) & \textbf{2.53} (0.67) \\
$n = 500$, $T = 50$ & 4.37 (0.65) & 2.86 (1.07) & 2.91 (2.01) & \textbf{0.99} (0.09) & 1.17 (0.18) \\
$n = 1000$, $T = 100$ & 4.14 (0.20) & 3.26 (1.21) & \textbf{0.76} (0.02) & 0.82 (0.03) & 1.06 (0.12) \\
$n = 2000$, $T = 200$ & 4.11 (0.12) & \textbf{0.72} (0.01) & \textbf{0.72} (0.01) & 0.76 (0.01) & 1.08 (0.12) \\
$n = 5000$, $T = 200$ & 4.11 (0.10) & 0.72 (0.01) & \textbf{0.71} (0.01) & 0.75 (0.02) & 1.37 (0.16) \\
\bottomrule
\end{tabular}
\end{table}

\begin{table}[!ht]
\centering
\caption{Simulation D: CRPS and Grassmann distances, mean~(sd) over 100 replications. Best per row is bold.}
\label{tab:simD_metrics}
\small
\setlength{\tabcolsep}{4pt}
\begin{tabular}{l l rrrrr}
\toprule
Setting $(n,T)$ & Metric & EOF & SCIEOF-RBF & SCIEOF-R2 & MESA Air & UK  \\
\midrule
\multirow{3}{*}{$(100, 20)$}
  & CRPS         & 2.85(1.15) & 2.91(1.30) & 3.21(1.32) & \textbf{1.44}(0.22) & 1.77(0.40) \\
  & $d_G^{(s)}$  & 1.29(0.16) & 0.42(0.29) & 0.86(0.46) & 0.26(0.04) & \textbf{0.16}(0.02) \\
  & $d_G^{(t)}$  & 1.64(0.04) & 1.37(0.10) & 1.55(0.19) & \textbf{0.22}(0.10) & 0.36(0.11) \\[4pt]
\multirow{3}{*}{$(500, 50)$}
  & CRPS         & 2.42(0.22) & 1.53(0.56) & 1.54(1.11) & 1.11(0.04) & \textbf{0.82}(0.09) \\
  & $d_G^{(s)}$  & 1.29(0.13) & \textbf{0.07}(0.03) & 0.20(0.32) & 0.10(0.01) & 0.09(0.00) \\
  & $d_G^{(t)}$  & 1.66(0.02) & 0.72(0.16) & 0.74(0.39) & \textbf{0.09}(0.03) & 0.12(0.03) \\[4pt]
\multirow{3}{*}{$(1000, 100)$}
  & CRPS         & 2.34(0.11) & 1.80(0.67) & \textbf{0.46}(0.01) & 1.16(0.12) & 0.74(0.06) \\
  & $d_G^{(s)}$  & 1.34(0.13) & 0.14(0.22) & \textbf{0.04}(0.00) & 0.06(0.00) & 0.06(0.00) \\
  & $d_G^{(t)}$  & 1.66(0.02) & 0.92(0.36) & 0.08(0.02) & \textbf{0.07}(0.02) & 0.09(0.02) \\[4pt]
\multirow{3}{*}{$(2000, 200)$}
  & CRPS         & 2.33(0.07) & 0.44(0.00) & \textbf{0.44}(0.00) & 1.23(0.08) & 0.73(0.05) \\
  & $d_G^{(s)}$  & 1.39(0.11) & \textbf{0.02}(0.00) & 0.03(0.00) & 0.04(0.00) & 0.04(0.00) \\
  & $d_G^{(t)}$  & 1.65(0.02) & \textbf{0.04}(0.02) & \textbf{0.04}(0.02) & 0.07(0.01) & 0.08(0.02) \\[4pt]
\multirow{3}{*}{$(5000, 200)$}
  & CRPS         & 2.33(0.06) & 0.44(0.00) & \textbf{0.44}(0.00) & 1.27(0.12) & 0.86(0.08) \\
  & $d_G^{(s)}$  & 1.39(0.10) & \textbf{0.02}(0.00) & \textbf{0.02}(0.00) & 0.04(0.00) & 0.04(0.00) \\
  & $d_G^{(t)}$  & 1.65(0.01) & 0.03(0.01) & \textbf{0.03}(0.01) & 0.06(0.01) & 0.11(0.02) \\
\bottomrule
\end{tabular}
\end{table}																																												Tables~\ref{tab:simD_rpmse}--\ref{tab:simD_metrics} show that, in the smallest sample size ($n$=100),  UK achieves the lowest RPMSE and best spatial subspace recovery, while MESA Air obtains the best CRPS and temporal subspace recovery, suggesting that its global smoothing strategy remains effective when the degree of nonstationarity is still limited.

As the sample size increases, SCIEOF-R2 and SCIEOF-RBF  emerge as the top two methods for subspace recovery at $n \ge 2000$, with substantial gains in both prediction and subspace recovery, while MESA Air also improves substantially and remains competitive in prediction. In particular, SCIEOF-R2 is able to recover the underlying spatial and temporal structure  with high accuracy  at moderate-to-large sample sizes. In contrast,  UK improves substantially through $n=2000$ but deteriorates at $n=5000$, while standard EOF remains largely unchanged, highlighting its inability to adapt to nonstationary, covariate-driven loading structures.

\subsection{Summary of Findings}\label{sec:sim_summary}

Across Simulations~A--D, the relative performance of the competing methods is largely determined by the agreement between their modeling assumptions and the underlying data-generating process.  Overall, sample size governs a consistent trade-off between estimation stability and model flexibility: when both $n$ and $T$ are small, UK and MESA Air often provide the most reliable predictions because their stronger structural assumptions reduce estimation variance,   whereas SCIEOF-R2 and SCIEOF-RBF may be unstable, since the   basis functions introduce many coefficients that cannot be estimated reliably from limited data.  This advantage gradually disappears as the sample size increases.  At larger sample sizes (e.g., $n \ge 1000$ or $n \ge 2000$), SCIEOF-R2  generally becomes one of the best-performing methods across all four simulations. Its flexible basis representation enables it to accommodate nonlinear spatial covariates, abrupt temporal variation, strong low-rank mean structures, and spatial nonstationarity without requiring these features to be specified through a parametric covariance model.

These patterns also point to a consistent limitation for each competing approach: EOF's purely unsupervised loadings cannot exploit covariate information; MESA Air's global smoothing can be less effective when abrupt or complex covariate-driven structure must be represented explicitly; and UK degrades once nonlinear mean structures or nonstationary dependence dominate over its assumed stationary covariance.

Finally, the results show that accurate subspace recovery alone does not necessarily guarantee accurate prediction. In Simulation~C, all methods recover similar latent subspaces despite exhibiting markedly different RPMSE and CRPS values, indicating that prediction accuracy depends  not only on subspace recovery but also on estimating the associated scores and signal magnitudes.

\section{Application to MERRA-2 Near-Surface Air Temperature}\label{sec:realdata}

\subsection{Data Description and Covariates}

Near-surface air temperature provides a useful application for evaluating covariate-informed latent-factor models because its variability reflects the combined influence of multiple processes acting across different spatial and temporal scales. In addition to large-scale atmospheric circulation, local terrain, boundary-layer dynamics, atmospheric moisture, cloud cover, precipitation, and the surface radiation budget all contribute to regional temperature variability. Many of these quantities are available directly in modern reanalysis products rather than remaining entirely latent.

The Modern-Era Retrospective Analysis for Research and Applications, Version~2 (MERRA-2), is a suitable data source for evaluating covariate-informed latent-factor models because it contains globally gridded atmospheric fields together with a broad collection of accompanying meteorological and land-surface variables. These variables are available on the same spatial and temporal grid as the response field, allowing observed environmental information to be incorporated directly into the low-rank decomposition. 

We analyze the near-surface air temperature field (T2M) from MERRA-2 at $0.5^\circ \times 0.625^\circ$ spatial resolution. The response variable is T2M in Kelvin without an anomaly transformation. The data were obtained from the surface-diagnostics, surface-flux, and radiation collections (M2T1NXSLV, M2T1NXFLX, and M2T1NXRAD),  and subsampled to 3-hourly resolution over 1–30 January 2020, yielding
\[
T = 30 \times 8 = 240
\]
time points. 
 The time-invariant constants were taken from M2C0NXASM.  We restrict the analysis to terrestrial grid cells with land fraction greater than $0.5$, resulting in $46{,}223$ candidate locations. To allow the largest sample size considered in the experiments, $n=50{,}000$, we additionally include cells with land fraction greater than $0.1$, yielding $51{,}924$ available locations. Thus the $n=50{,}000$ experiment is not drawn from the same spatial population as the smaller designs; it additionally includes mixed land–water cells.

T2M variability is associated with a range of observed atmospheric and surface conditions. To examine how these variables affect the estimated latent structure, we include three groups of covariates. The first group consists of $K=6$ \emph{spatio-temporal} covariates, which enter the additive component
$\sum_{k=1}^{K} b_k \bm{F}_{(k)}$ in equation~(\ref{eq:SVD_model}). Each variable is converted into a location-specific temporal anomaly before analysis:

\begin{itemize}
\item \textbf{PBLH} (planetary boundary layer height) and \textbf{PRECTOTCORR} (bias-corrected total precipitation) from M2T1NXFLX;
\item \textbf{Wind speed} $\sqrt{U_{10}^2 + V_{10}^2}$ and \textbf{QV2M} (2-m specific humidity) from M2T1NXSLV;
\item \textbf{SWGDN} (surface downward shortwave flux) and \textbf{CLDTOT} (total cloud fraction) from M2T1NXRAD.
\end{itemize}

For the second group, five \emph{spatial-only} covariates form the columns of $\bm{R}$: elevation (PHIS$/g$), absolute latitude $|\mathrm{lat}|/90^\circ$, land fraction, $\log(1+\mathrm{SGH})$ (sub-grid orography), and great-circle distance to the nearest coastline. For the third group, four \emph{temporal-only} covariates form  the columns of  $\bm{H}$: the diurnal harmonics $\cos(2\pi h/24)$ and $\sin(2\pi h/24)$ of UTC hour, 
together with the daily Arctic Oscillation (AO) and North Atlantic Oscillation (NAO)
 indices. The daily AO and NAO values are held constant across the eight 3-hourly steps within each calendar day.

\subsection{Implementation and Evaluation Setup}

 The complete MERRA-2 land grid contains tens of thousands of spatial locations, making full-data evaluation   computationally challenging, particularly for covariance-based methods. To ensure a fair comparison, we draw uniform random subsamples of terrestrial cells with sample sizes $n\in\{500,\,2000,\,5000,\,10\,000,\,20\,000,\,40\,000,\,50\,000\}$ 
and evaluate all methods on the same subsamples for each size. Each subsample is split
with the same scheme as the simulations into training and evaluation sets by holding out 20\% of spatial locations and 20\% of time points. Reported scores pool all held-out space–time entries under that scheme.

Because the domain is the Earth's surface rather than a planar region, 
each cell's spatial coordinate is mapped to a 3-D unit vector $\bm{s} = (\cos\gamma\cos\eta,\ \cos\gamma\sin\eta,\ \sin\gamma)$, where $\gamma$ and $\eta$ denote latitude and longitude, respectively, so that Euclidean distance between cells equals chord distance on the sphere. 

The MRTS-based method here, denoted \emph{SCIEOF-SphMRTS}, replaces the planar MRTS of SCIEOF-R2 with its spherical counterpart, evaluating the thin-plate kernel directly on $S^2$ \citep{huang2026multi}; the two are otherwise identical. 

All competing methods use the same covariate information. The SCIEOF variants and Standard EOF  incorporate the spatial-only and temporal-only covariates through columns of $\bm{R}$ and $\bm{H}$, respectively, while the spatio-temporal covariates enter in the additive  terms. UK includes the covariates as fixed-effect drift terms, whereas MESA Air uses the spatio-temporal and temporal covariates as fixed effects and the spatial covariates as predictors in its land-use regression component. 
UK uses the product-sum  space-time Vecchia approximation with $m = 40$ nearest neighbours, fit by maximum likelihood on a 1500-point subset.

The basis specifications differ across the SCIEOF variants. Let $p^*=p-p_0$ and $q^*=q-q_0$ denote the numbers of spatial and temporal basis functions, respectively, with cubic B-splines used for the temporal basis. For SCIEOF-SphMRTS, when $n\le 2000$, we use the default specification $p^*=\min(200,n-5)$ and $q^*=17$. For $n\in\{5000,10\,000\}$, we increase the spatial basis dimension to $p^*=800$ and use $q^*=60$ temporal basis functions. For $n\ge20\,000$, we further increase the spatial basis dimension to $p^*=2500$, while retaining $q^*=60$.

SCIEOF-RBF uses a different spatial basis and a different choice of $p^\ast$.  The spatial basis consists of single-scale Gaussian kernels in chordal distance,
$\phi_j(\bm{s})
=
\exp\left\{
-d^2(\bm{s},\bm{s}_j)/h_n^2
\right\}$ with
$h_n=\frac{2}{\ell-1}$, where $\ell=\sqrt{p^*}$ determines the resolution along each coordinate direction (longitude and latitude), with
$\ell
=
\max\left\{
4,\,
\operatorname{round}\left(4(n/100)^{1/4}\right)
\right\}$.
The centers are placed at a fixed uniform random subsample of the training sites. At $n=5000$, $p^*=121$ (so $\ell=11$), with an e-folding angular radius of approximately $10^\circ$.

Although $T=240$ is common across all sample sizes, the usable basis dimension is constrained by the number of observed sites. In particular, the bilinear factorization estimates $O\!\bigl(L(p^*+q^*)\bigr)$ parameters from $n_{\mathrm{obs}}\times T_{\mathrm{obs}}$ observations. During preliminary tuning, we observed that richer basis specifications for $n\le2000$ could cause the internal rank selection to collapse to $L=2$, with degraded predictive performance. We therefore retain the smaller default basis for these sample sizes and use the tuned $q^*=60$ specification from $n=5000$ onward.

\subsection{Results}

Tables~\ref{tab:t2m_rpmse} and~\ref{tab:t2m_crps} report RPMSE and CRPS for the T2M response (in K) over the held-out test set at seven subsample sizes. Best per column is bold. SCIEOF-SphMRTS uses the spherical MRTS basis with the rank selected by cross-validation. 

\begin{table}[!ht]
\centering
\caption{ Comparison of RPMSE for raw 3-hourly global T2M (January 2020) on the held-out test set across different subsample sizes. $\dagger$~MESA Air at $n = 50\,000$ exceeded available memory (its per-coefficient-field REML kriging requires an $n \times n$ factorization) and was terminated.}
\label{tab:t2m_rpmse}
\small
\setlength{\tabcolsep}{3pt}
\begin{tabular}{l r r r r r r r}
\toprule
Method & $n=500$ & $n=2000$ & $n=5000$ & $n=10\,000$ & $n=20\,000$ & $n=40\,000$ & $n=50\,000$ \\
\midrule
EOF                  & 4.19 & 4.11 & 4.01 & 3.92 & 4.05 & 3.85 & 3.78 \\
SCIEOF-RBF           & 4.82 & 6.42 & 4.47 & 4.32 & 4.23 & 4.70 & 6.17 \\
SCIEOF-SphMRTS       & 4.26 & 3.24 & 1.85 & 1.81 & \textbf{1.47} & \textbf{1.49} & \textbf{1.58} \\
MESA Air             & 3.42 & 3.26 & 3.08 & 3.03 & 3.03 & 2.94 & ---$^{\dagger}$ \\
UK    & \textbf{2.58} & \textbf{2.08} & \textbf{1.75} & \textbf{1.65} & 1.69 & 1.59 & 1.59 \\
\bottomrule
\end{tabular}
\end{table}

\begin{table}[!ht]
\centering
\caption {Comparison of CRPS for raw 3-hourly global T2M (January 2020) on the held-out test set across different subsample sizes. $\dagger$~MESA Air at $n = 50\,000$ exceeded available memory (its per-coefficient-field REML kriging requires an $n \times n$ factorization) and was terminated.}
\label{tab:t2m_crps}
\small
\setlength{\tabcolsep}{3pt}
\begin{tabular}{l r r r r r r r}
\toprule
Method & $n=500$ & $n=2000$ & $n=5000$ & $n=10\,000$ & $n=20\,000$ & $n=40\,000$ & $n=50\,000$ \\
\midrule
EOF                  & 2.30 & 2.26 & 2.21 & 2.16 & 2.14 & 2.12 & 2.08 \\
SCIEOF-RBF            & 2.68 & 3.59 & 2.47 & 2.38 & 2.34 & 2.60 & 3.43 \\
SCIEOF-SphMRTS        & 2.33 & 1.80 & \textbf{1.00} & \textbf{0.98} & \textbf{0.81} & \textbf{0.81} & \textbf{0.86} \\
MESA Air             & 7.64 & 10.30 & 10.44 & 11.63 & 11.27 & 12.41 & ---$^{\dagger}$ \\
UK    & \textbf{1.80} & \textbf{1.44} & 1.20 & 1.13 & 1.15 & 1.02 & 1.03 \\
\bottomrule
\end{tabular}
\end{table}

The results here broadly reinforce the conclusions from the simulation study, while also illustrating how the competing methods behave for a much larger spatio-temporal field on the sphere. Standard EOF again shows only modest improvement as the number of spatial locations increases, suggesting that estimating latent modes solely from the empirical covariance gains relatively little from additional observations. SCIEOF-RBF likewise exhibits inconsistent performance, indicating that localized radial basis functions are less
effective for representing the dominant large-scale temperature patterns in MERRA-2.

MESA Air maintains relatively stable RPMSE across all sample sizes, but its CRPS increases noticeably as $n$ grows. A possible explanation is that the globally smoothed temporal representation and  restrictive covariance structure become   less capable of capturing the heterogeneous uncertainty  in the large-scale temperature field, so that point predictions remain serviceable while the predictive distributions are substantially miscalibrated.

UK continues to perform competitively for small and moderate sample sizes, although its improvement becomes negligible beyond approximately $10{,}000$ locations. In contrast, SCIEOF-SphMRTS shows the largest gains as $n$ increases, with sharp declines in RPMSE and CRPS from $n = 500$ to $n = 20,000$, before its performance begins to plateau. Although the improvement plateaus beyond $n = 20,000$, SCIEOF-SphMRTS achieves the lowest  CRPS among all methods for $n \ge 5,000$, and the lowest RPMSE once $n\ge 20{,}000$. At $n=5{,}000$ and $n=10{,}000$, UK remains slightly better in RPMSE.   Although these trends are consistent with the simulation study, they are particularly encouraging for this real application, demonstrating that the proposed method remains effective for modeling large-scale temperature fields on the sphere while retaining computational scalability.

The six spatio-temporal covariates enter as additive fields $\sum_{k=1}^{K}b_k f_k (\bm{s},t)$. Table~\ref{tab:st_coef} reports the estimated coefficients at the $n=5{,}000$ fit used for the modal maps below and at the largest design $n=50{,}000$. At both sample sizes the associations with T2M are positive for planetary-boundary-layer height, precipitation, 2-m specific humidity, downward shortwave radiation, and total cloud fraction, and negative for 10-m wind speed. Because these terms have been partialled out, the leading SCIEOF components in Figures~1 and~2 represent residual large-scale structure rather than the local radiative and boundary-layer effects already captured by the covariates.

\begin{figure}[!ht]
\centering
\includegraphics[width=\linewidth]{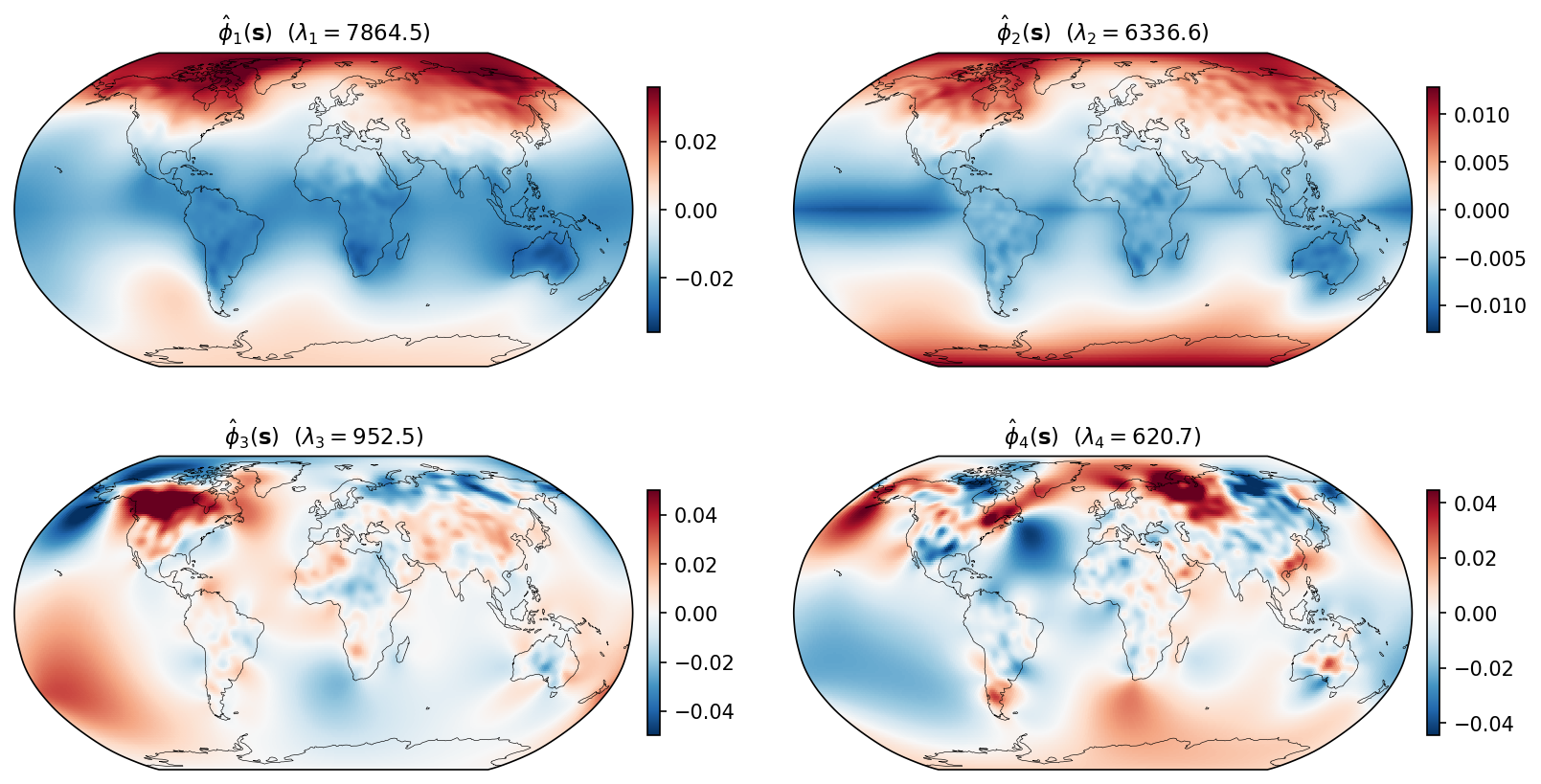}
\caption{Leading four estimated spatial loadings $\hat\phi_j(\bm{s})$ (Robinson projection) from the SCIEOF-SphMRTS fit to real-data T2M at $n = 5000$, ordered by singular value $\lambda_j$. Red/blue denote positive/negative loading.}
\label{fig:svd_spatial}
\end{figure}

\begin{figure}[!ht]
\centering
\includegraphics[width=0.9\linewidth]{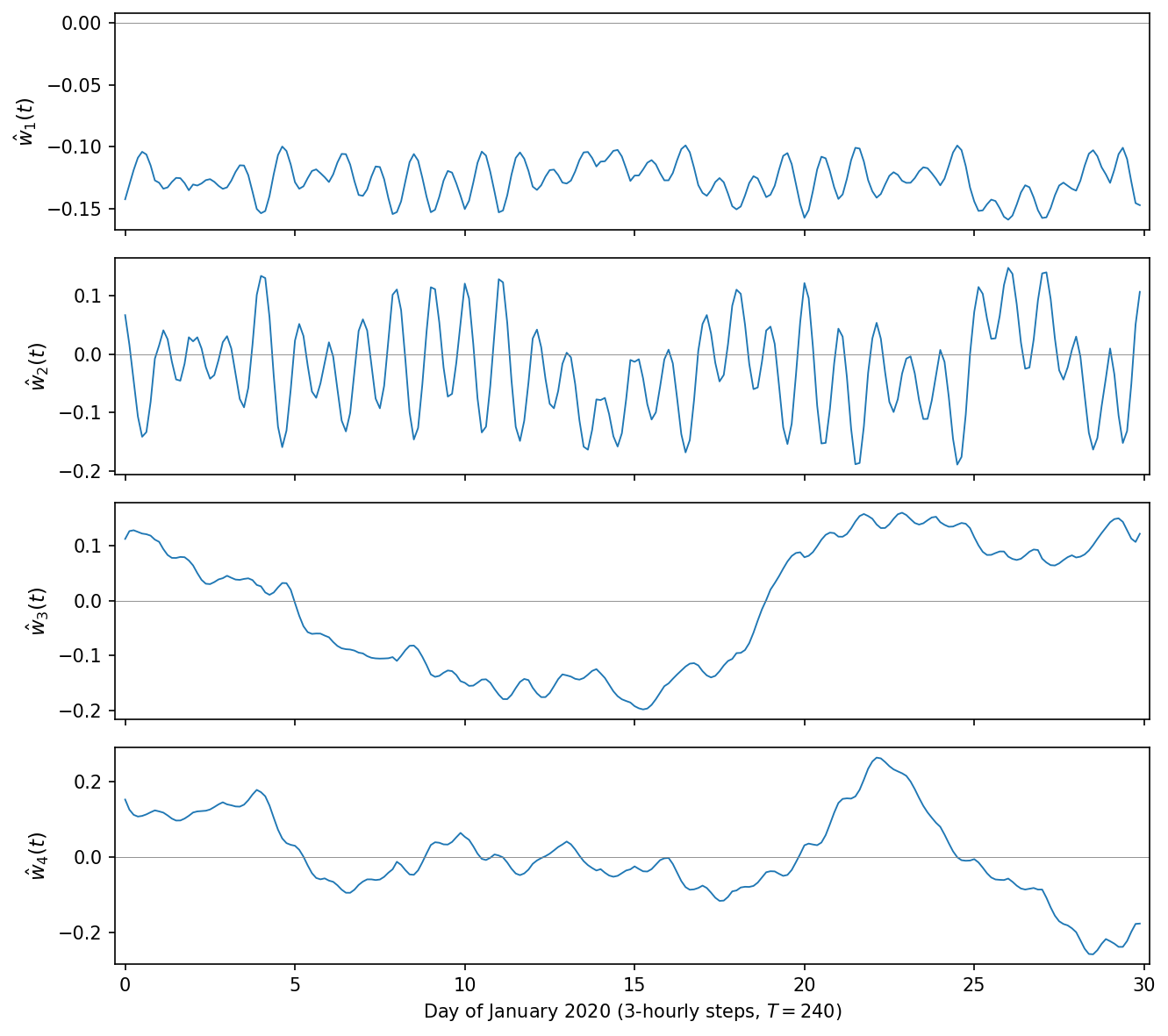}
\caption{Leading four estimated temporal scores $\hat w_j(t)$ from the same SCIEOF-SphMRTS fit, over the $T = 240$ three-hourly time steps of 1–30 January 2020. $\hat w_2$ shows the once-per-day diurnal oscillation; $\hat w_3$ and $\hat w_4$ capture synoptic sub-monthly variation.}
\label{fig:svd_temporal}
\end{figure}

Beyond predictive accuracy, a key advantage of the factorization in SCIEOF-SphMRTS is its ability to extract physically interpretable latent modes after controlling for explicit spatio-temporal covariates.
Figures~\ref{fig:svd_spatial} and~\ref{fig:svd_temporal} display the leading four  estimated spatial loadings $\hat{\phi}_j(\bm{s})$ and corresponding temporal scores $\hat{w}_j(t)$ fit on the global T2M dataset ($n = 5000$). The four
components carry singular values $(\lambda_{1},\ldots,\lambda_{4} ) = (7864.5,\, 6336.6,\, 952.5,\,
620.7)$, i.e.\ $58.2\%$, $37.8\%$, $0.9\%$, and $0.4\%$ of
$\sum_j \lambda_j^2$ over the $L = 64$ retained components, and together account for
$97.3\%$ of the covariate-adjusted variability in the reconstruction sense (the full
rank-64 fit captures $99.4\%$). 
Because explicit spatio-temporal covariates have been regressed out, these components represent the dominant residual latent structures.

Components 1 and 2 describe coherent climatological patterns over the $T=240$ three-hourly steps (1–30 January 2020). 
    The loading map $\hat{\phi}_1(\bm{s})$ captures a clear continental--ocean contrast, while $\hat{\phi}_2(\bm{s})$ exhibits a distinct hemispheric/latitudinal gradient (Northern-winter versus Southern-summer contrast). 
   Their corresponding scores ($\hat{w}_1, \hat{w}_2$) roughly capture  the baseline daily mean temperature and the localized timing (phase and amplitude) of the 24-hour diurnal cycle, respectively.  

    Components 3 and 4  separate high-frequency daily oscillations from sub-monthly, synoptic-scale weather variability.  
    The temporal scores $\hat{w}_3(t)$ and $\hat{w}_4(t)$ display low-frequency, multi-day shifts,    potentially associated with winter storm tracks and regional air-mass transitions. 
    Their spatial loadings ($\hat{\phi}_3, \hat{\phi}_4$) reveal prominent localized dipole structures, particularly across mid-to-high latitudes. This demonstrates SphMRTS's ability to capture fine-scale spatial heterogeneity without over-smoothing, reflecting large-scale atmospheric dynamic gradients.

\begin{table}[t]
\centering
\caption{Additive coefficients in native units (K per unit of the covariate)
for SCIEOF-SphMRTS at $n=5{,}000$ and $n=50{,}000$.}
\label{tab:st_coef}
\begin{tabular}{llrr}
\toprule
Covariate & Unit & $n=5{,}000$ & $n=50{,}000$ \\
\midrule
PBLH & m & 0.0008365 & 0.0008655 \\
PRECTOTCORR & kg\,m$^{-2}$s$^{-1}$ & 28.72 & 99.79 \\
WIND & m\,s$^{-1}$ & -0.03315 & -0.01799 \\
QV2M & kg\,kg$^{-1}$ & 238.6 & 136 \\
SWGDN & W\,m$^{-2}$ & 0.007845 & 0.007079 \\
CLDTOT & -- & 0.8941 & 0.6888 \\
\bottomrule
\end{tabular}
\end{table}

\section{Discussion}\label{sec:discuss}

This work introduces SCIEOF as a covariate-informed extension of classical EOF analysis that bridges unsupervised low-rank decomposition and prediction-oriented spatial modeling. Rather than estimating latent modes solely from the empirical covariance, SCIEOF allows known spatial-only and temporal-only covariates to directly shape the composition of the latent spatial and temporal modes, while the effect of spatio-temporal covariates is isolated through a separate additive term. The factorized components are thus bilaterally informed by external information. Specifically, spatial modes and temporal scores are estimated jointly with observed covariates rather than being determined solely by the covariance structure, and remain directly comparable to conventional EOF modes. In this way, SCIEOF complements covariance-based spatial models by allowing structural information and scientific knowledge to be encoded through the latent representation itself, rather than treated as a nuisance to be regressed out beforehand.

The simulation study and the MERRA-2 application demonstrate several practical advantages of this formulation. When informative covariates are available, explicitly modeling their contributions improves predictive performance while producing a more parsimonious latent representation. Moreover, the real-data analysis shows that SCIEOF-SphMRTS continues to benefit from increasing spatial sample sizes, whereas several competing methods exhibit only marginal improvement or become computationally restrictive. These results suggest that incorporating appropriate structural information can be as important as increasing the number of observations when analyzing large spatio-temporal environmental datasets.

The proposed framework nevertheless has several limitations. The decomposition depends on both the selected covariates and the chosen spatial basis. If important explanatory variables are omitted, their effects remain embedded in the latent factors, whereas highly correlated covariates may influence how variability is partitioned between the regression component and the residual low-rank structure. Although the current two-stage estimation procedure performs well in both simulations and real applications, jointly estimating the regression and latent components may further improve the separation between explained and latent variability.

More broadly, SCIEOF provides a general framework for introducing structural information into latent-factor models. In this work, the structural information is represented through linear spatial, temporal, and spatio-temporal covariates. Future research may extend this idea by incorporating richer forms of prior knowledge. For example, physical constraints derived from conservation laws or advection--diffusion equations could be integrated directly into the latent score dynamics, allowing the decomposition to respect known transport mechanisms while maintaining a low-dimensional representation. Likewise, these structural components may serve as inductive biases within modern machine learning frameworks, such as physics-informed neural networks or other deep generative models, combining the interpretability of structured latent factors with the flexibility of non-linear representations. Such extensions would broaden the applicability of SCIEOF to increasingly complex environmental systems while preserving its central objective of separating known structure from genuinely latent variability.



\section*{Acknowledgements}
This work was partially supported by the National Science and Technology Council, Taiwan, under Grant Nos. NSTC115-2111-M008-035 and NSTC115-2118-M259-003. 

\bibliographystyle{plainnat}
\bibliography{reference}

\end{document}